\documentclass[11pt,twoside]{article}

\usepackage{asp2006}
\usepackage{epsf}
\usepackage{graphicx}
\usepackage{lscape}
\usepackage{sidecap}

\begin{document}
\title{X-ray absorption in Active Galactic Nuclei}   
\author{Roberto Maiolino\altaffilmark{1} and Guido Risaliti\altaffilmark{2}}   
\altaffiltext{1}{INAF -- Astronomical Observatory of Rome, Italy}
\altaffiltext{2}{INAF -- Arcetri Astrophysical Observatory, Firenze, Italy}

\begin{abstract} 
We review some of the main physical and statistical properties of the X-ray 
absorber in AGNs. In particular, we review the
distribution of the absorbing column density inferred from X-ray observations of
various AGN samples. We discuss the location of the X-ray absorber and the relation
with the dust absorption at optical and infrared wavelengths. Finally, we
shortly review the recent findings on X-ray absorption at high luminosities and
at high redshift.
\end{abstract}


\section{Introduction}   

The X-ray absorption provides important information on the
nature of the circumnuclear medium in AGNs.
Understanding the physical and statistical properties of the X-ray absorption
is also highly relevant to characterize the black hole accretion
in the universe. Indeed, most of the AGN activity, both locally
and at high redshift, is obscured.
In this short paper we shortly review the main observational results in this field.
Since absorption by ``warm'' gas is also treated
by other reviews in these proceedings, ``warm absorption'' is only shortly
discussed in Sect.2. We will then focus on the ``cold absorption'', by discussing
it statistical and physical properties, as well as some of the recent findings at high redshift.
It is important to clarify that,
due to the limited space available, this review in unavoidably highly incomplete.

\section{The warm absorber}

%
At least 50\% of the low resolution X-ray spectra of type 1 (optically unobscured)
AGNs show the presence of a broad absorption feature at $\sim$0.7--0.8~keV, ascribed
to absorption edges of OVIII and OVII due to a ``warm'', ionized gas
along our line of sight \citep{reynolds97,crenshaw99}. However, the
advent of grating X-ray spectroscopy with Chandra and XMM improved dramatically our
understanding of warm absorbers.
In particular, the recent, high resolution X-ray spectra revealed that the
trough at $\sim$0.7--0.8~keV
is actually the blend of various absorption lines and absorption edges.
The 900~ks Chandra HETGS spectrum of NGC~3783 is
the high resolution spectrum with the highest quality currently available,
and it has been extensively
used to investigate the properties of the warm absorber
\citep{kaspi02,netzer03,krongold03}. It was found that, at least in the case of
NGC~3783, the warm absorber has a column density of the order of a few
times $\rm 10^{22}~cm^{-2}$, it is outflowing with velocities ranging from a few
to several $\rm 100~km~s^{-1}$ and it is located between $\sim$0.2~pc and $\sim$3~pc
from the ionizing, nuclear source. According to \cite{netzer03} and \cite{krongold03}
matching the several absorption lines observed in the spectrum of NGC~3783 requires
the presence of two or three different phases of the absorbing medium, at different
temperatures and different ionization stage, but in pressure equilibrium. However,
\cite{gonclaves06} pointed out that multi-temperature components arise naturally in a
single medium as a consequence of the stratification
of the ionization structure of each cloud exposed to the nuclear source. In
particular, \cite{gonclaves06} could fit the various absorption lines observed in
NGC~3783 with a single medium, without the need for different, separate components of
the absorbing medium.

\section{Cold absorption: general properties and warnings}

The presence of a cold, neutral medium along the line of sight introduces a sharp
photoelectric absorption cutoff in the power-law spectrum emitted by the nuclear
source. It is possible to accurately
determine the column of absorbing material by measuring
the energy of the photoelectric absorption. At
$\rm N_H > 10^{24}~cm^{-2}$ the gas is thick to Compton scattering (dubbed as
``Compton thick''). In this case the primary X-ray radiation is
completely absorbed at energies $<$10~keV. However, as long as the column
of gas does not exceed $\rm 10^{25}~cm^{-2}$ the primary radiation is still
transmitted and observable at energies in the range 10--100~keV. In Compton thick
sources with $\rm N_{H}> 10^{25}~cm^{-2}$ the direct, primary X-ray
radiation is totally absorbed at any energy \citep{matt97}.

Although, the primary radiation is totally absorbed (at least at $\rm E<10~keV$),
Compton thick sources are still observable through radiation that is scattered into
our line of sight either by a cold, Compton thick medium (``cold reflection'')
 or, less frequently, by a warm medium (``warm
reflection''),
either of such scattering media must extend on scales larger than the absorber.
The reflected component is about two orders of magnitude fainter than
the primary radiation; therefore Compton thick sources are
much more difficult to detect, especially at high redshift.
Compton thick, reflection-dominated sources are generally characterized also by the
presence of a prominent FeK$\alpha$ line at 6.4~keV. This line is partly
produced in the accretion disk and is partly excited in the circumnuclear medium on
larger scales. In Compton thin sources this iron line is heavily diluted by the
direct, primary radiation, and its observed equivalent width is of a few hundred eV.
In Compton thick sources the primary continuum is suppressed, and therefore
the iron line emitted by the circumnuclear medium is observed with an equivalent width
which easily exceeds 1~keV.

Note however that an X-ray spectrum which appear reflection-dominated,
and with a prominent iron K$\alpha$ line, does not necessarily imply that
the source is Compton thick along our line of sight. Indeed, if the active nucleus
fades, its light echo keeps the circumnuclear
medium reflecting the radiation for several years, producing a reflection-dominated
spectrum, even if the nucleus is totally unobscured.
Compton thin to Compton thick transitions have been sometimes
interpreted in terms of this ``fossil'' scenario \citep{guainazzi02,matt03}. However,
in most of these cases it is difficult to distinguish whether the spectral change
is really due to an intrinsic fading of the source or to an increase of the absorbing
column density (see Sect.5). Yet, in a few cases the systemic decline of the
luminosity, monitored through various epochs, unambiguously identifies the Compton
thick--like appearance of the final spectrum as due to the fossil nature of the
source \citep{gilli00}.

Finally, it should be noted that
absorption in the hard X-rays is due to metals, therefore what we actually
is the column of metals. To infer the equivalent column of
hydrogen people generally assume (explicitly or, more often, implicitly)
solar abundances. However, nearly all AGNs display super-solar abundances
\citep{hamann02,nagao06a,nagao06b}. As a consequence, hydrogen column
densities inferred from X-ray spectra are generally overestimated.

\section{The $\rm N_H$ distribution among AGNs}

\begin{SCfigure}
\includegraphics[width=0.45\textwidth]{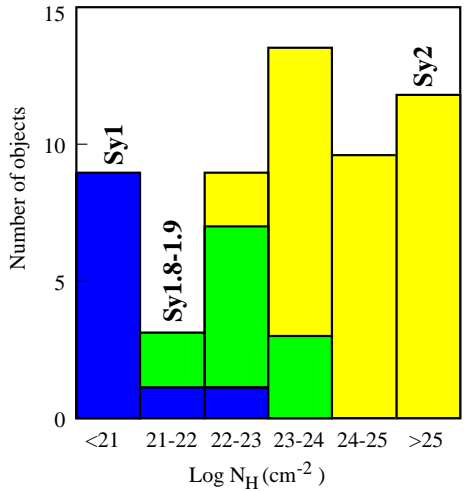}
\caption{$\rm N_H$ distribution for AGN selected in [OIII]5007 line flux,
divided in different optical types.
Taken from \cite{salvati00} \citep[adapted from][]{risaliti99}.}
\label{nh_dist}
\end{SCfigure}

Early investigations with hard X-ray satellites clearly revealed an excess of
absorption in type 2 Seyferts, in agreement with the expectations
from the unified model. However, such early studies
could identify only very few Compton thick sources, suggesting that the latter
is a very rare class of objects. Yet, later, deeper surveys, adopting careful selection
criteria and exploiting a wider energy range, discovered a much larger fraction of
Compton thick AGNs \citep{maiolino98,bassani99}. By extracting a subsample selected
in [OIII]$\lambda$5007 flux, assumed to trace the intrinsic AGN flux,
\cite{risaliti99} could determine a first, unbiased distribution of $\rm N_H$ among
Seyfert nuclei (Fig.\ref{nh_dist}). Among optically obscured Seyfert nuclei the distribution
of $\rm N_H$ is nearly flat, and Compton thick Seyfert 2s
($\rm N_H > 10^{24}~cm^{-2}$) are found to be as numerous as Compton thin ones.
Generally there is a good correspondence between optical classification and X-ray
absorption: Sy1s tend to have little or no absorption, ``strict'' type 2 Seyferts tend
to be heavily absorbed ($\rm N_H > 10^{23}~cm^{-2}$),
while intermediate type 1.8--1.9 Seyferts are absorbed
by intermediate $\rm N_H$ ($\sim 10^{22}-10^{23}~cm^{-2}$).

The column density distribution has been extracted also for other samples of AGNs selected
in different ways. \cite{markwardt05} and \cite{bassani06} give the $\rm N_H$
distribution of AGNs selected in the 10--100~keV energy range. These samples are
less biased against obscured AGNs, with respect to samples selected at lower energies, but
they are still biased against Compton thick AGNs. Indeed, as discussed above,
AGNs with $\rm N_H > 10^{25}~cm^{-2}$ are absorbed at all energies, while partially
Compton thick AGNs with $\rm 10^{24} < N_H < 10^{25}~cm^{-2}$ do show a transmitted
component at E$>$10~keV, but still significantly absorbed with respect to Compton thin
AGNs. Indeed, the samples presented by \cite{markwardt05} and \cite{bassani06}
show the absence of AGNs with $\rm N_H > 10^{25}~cm^{-2}$ and a paucity of AGNs with
$\rm 10^{24} < N_H < 10^{25}~cm^{-2}$.

Radio emission is another selection which should be free of absorption
biases. Indeed,
early X-ray studies of radio-loud AGNs revealed a large fraction of obscured
AGNs \citep{sambruna99}. The absorbing column density is also found to anti-correlate
with the radio core dominance parameter R (a measure of the jet orientation),
in agreement with the expectations from the unified model
\citep{grandi06}. However, there is a puzzling shortage of Compton thick AGNs, which is
confirmed also in more recent studies \citep{evans06,hardcastle06}.
A possible explanation is that the radio jet
contributes (or dominates) the X-ray luminosity, which is therefore inefficiently obscured
by a compact, circumnuclear medium. 

Recently, \cite{cappi06} measured the column density with XMM in a distance
limited sample of Seyferts 
\citep[pre-selected through optical spectroscopy,][]{ho97}.
Within the statistical uncertainties, the resulting $\rm N_H$
distribution is similar to
that obtained in previous optically selected samples.

One of the major limitations of the previous surveys is that in most of them AGNs were
pre-selected to have a Seyfert-like optical spectrum. However, hard X-ray observations
\citep{vignati99,guainazzi00,dellaceca02}
have revealed the presence of heavily obscured, relatively luminous Seyfert nuclei
($\rm L_{2-10~keV}>10^{42}~erg~s^{-1}$) in galaxies
optically classified as starburst (HII)
or LINER (here with the latter term we refer to emission due to shocks in
starburst superwinds, not nuclear LINERs). Similar results have been obtained
through near- and mid-IR spectroscopy \citep{imanishi06,risaliti06}.
\cite{maiolino03} investigated the statistical properties of such optically elusive
Seyfert nuclei, and found that most of them are Compton thick. Therefore, the actual
fraction of Compton thick AGNs is higher than inferred in samples where objects are
pre-selected through the optical spectrum.
Note that evidence for relatively powerful Seyfert nuclei hosted in
galaxies which are apparently normal in the optical has also been found in high
redshift surveys \citep[e.g.][]{comastri02,szokoly04,barger05,cocchia07}. However,
in several of these cases the optical mis-classification may simply be due to
dilution of the nuclear light by the host galaxy \citep{moran02} or to inappropriate
(rest-frame) spectral coverage \citep{severgnini03}.

\section{The location of the cold X-ray absorber}

Location, size and geometry of the absorbing medium have been some
of the most debated topics.
Although somewhat artificially, the gaseous, X-ray absorber can be roughly divided in two main
components: an extended medium (on scales of 10--100~pc) and a compact, nuclear
absorber ($<$1--10~pc).

Evidence for an absorbing medium on scales of about 100~pc, possibly associated with the
host galaxy gaseous disk, has been inferred from the statistical properties
of the hosts and of the NLR of Seyfert galaxies \citep[e.g.][]{maiolino95}, from the
direct observation of obscuring structures in high resolution images
\citep[e.g.][]{malkan98} and from the detection of circumnuclear molecular gas
\citep[e.g.][]{schinnerer99}. The X-ray absorption properties of such large
scale absorbers have been studied with some detail by \cite{guainazzi05}.
They found that Seyfert nuclei with dusty structures, observed within the central
few 100~pc in HST images, are typically characterized by an absorbed, but Compton thin
X-ray spectrum.

\begin{SCfigure}
\includegraphics[width=0.45\textwidth]{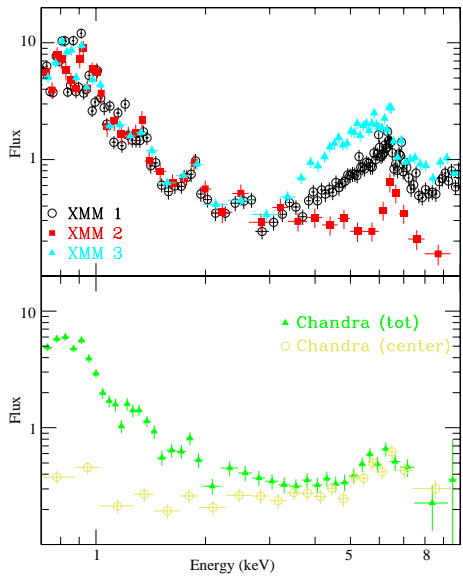}
\caption{Unfolded spectra of NGC~1365 taken during a period of about 8 months showing strong
$\rm N_H$ variations \citep[from][]{risaliti05}.
The most rapid variation is observed between observations XMM1 and XMM2,
which are only three weeks apart, and which show a transition from Compton thin to Compton thick.
}
\label{nh_var}
\end{SCfigure}

Evidence for an additional, much more compact gaseous absorber is directly
obtained from X-ray data. There are two lines of evidence, one based on dynamical mass
constraints and the other one on time variability.

\cite{risaliti99} showed that, for any reasonable geometry, gas with
$\rm N_H > 10^{24}~cm^{-2}$ cannot be accommodated on scales larger than a few 10~pc
without exceeding the dynamical mass in the same region. Therefore, the Compton thick
medium must be located within the central $\sim$10~pc.

Tighter constraints come from the observed temporal variability of the absorbing column
density. \cite{risaliti02} showed that variability of the X-ray absorbing column density
on time scales of years is observed in nearly all AGNs for which multi-epoch X-ray observations
are available.
For a subsample of the sources variability is observed even on scale shorter than one year.
For any reasonable geometry of the individual clouds, the latter result implies
that in these sources most of the the X-ray absorption must occur on sub-parsec scales.
Even tighter constraints come from dedicated X-ray monitoring of some individual sources
\citep{risaliti05,elvis04}.
These observations revealed strong $\rm N_H$ variations, by even passing from the Compton thin
to the Compton thick regime, on time scales of weeks or even as short as a few hours
(Fig.\ref{nh_var}). These
rapid variations indicate that the absorber must much closer to the source than the standard
pc-scale model, and probably co-spatial with the Broad Line Region. Further results
and details on the variability of X-ray absorption are given in the contribution by Risaliti
within these conference proceedings.

\section{X-ray absorption versus optical and infrared absorption}

Various studies have shown that X-ray and optical nuclear
absorption do not match in AGNs. In particular, the measured
optical dust extinction is
systematically lower than inferred from the column density $\rm N_H$ measured in
the X-rays, assuming a Galactic gas-to-dust ratio \citep{maiolino01a}.
An important consequence of this effect is the mismatch between optical and X-ray
classification, and in particular the existence of type 1, broad line AGN with significant
X-ray absorption \citep{wilkes02,hall06,szokoly04,barger05,fiore03,silverman05}.
Extreme cases of this kind of mismatch are Broad Absorption Line (BAL) QSOs whose X-ray
spectrum is heavily absorbed, and in some cases even Compton thick, although their optical
spectrum shows little or no dust extinction \citep{gallagher06,braito04,maiolino01c}.

For what concerns the origin the mismatch between X-ray and optical
absorption, there are two possible physical reasons.
The BLR is dust free, because inside
the dust sublimation radius; therefore, if
a large fraction of the X-ray absorbing column density is located within the BLR, as
discussed in the previous section, then this naturally yields to a reduced $\rm A_V/N_H$.
Additionally, the circumnuclear dusty medium of AGNs is very dense 
($\rm n\sim 10^{5}~cm^{-3}$) and in such dense environments dust grains tend to be
larger, therefore being less effective in absorbing the optical and UV radiation
\citep{maiolino01b}.

\begin{SCfigure}
\includegraphics[width=0.5\textwidth]{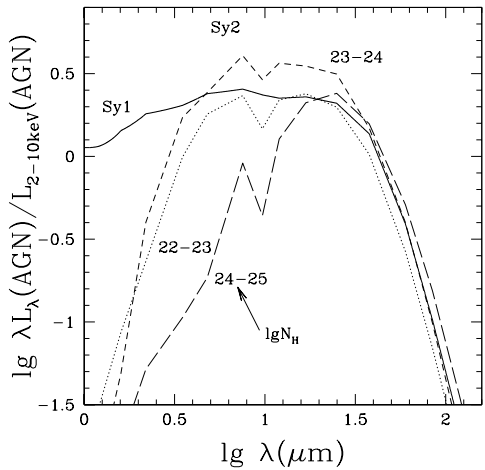}
\caption{Nuclear, infrared spectral energy distribution observed in local AGNs,
averaged in bins of absorbing $\rm N_H$ inferred from the X-rays. The IR SEDs
are normalized to the absorption-corrected, hard X-ray luminosity. There is little variation of
the nuclear IR SED for different values of $\rm N_H$, except for Compton thick sources, which
show a significant depression short-ward of 10$\mu$m.}
\label{nh_ir}
\end{SCfigure}

Similar results have been obtained from the comparison between infrared and X-ray
absorption. By using Spitzer mid-IR spectra, \cite{shi06} found that the depth
of the dust silicate feature at $\rm \sim 9.7~\mu m$ correlates with the X-ray absorption,
though with a large scatter. However, in most cases the silicate feature is
much shallower than expected from the $\rm N_H$ inferred from the X-rays by assuming a Galactic
gas-to-dust ratio.
Also the intensity and shape of the mid-IR continuum are little related by the presence
of X-ray absorption along the line of sight. More specifically, \cite{silva04} found
that the shape and intensity of nuclear mid-IR continuum of AGNs is essentially unchanged
and independent of the column density $\rm N_H$ measured in the X-rays (Fig.\ref{nh_ir}).
Only for Compton
thick sources the mid-IR SED appears reddened at wavelengths shorter than $\sim 10 \mu m$,
but the inferred extinction is still more than one order of magnitude lower than expected
from the $\rm N_H$ inferred from the X-rays. Similar results were obtained by \cite{lutz04}
and \cite{krabbe01}. The motivations for the reduced IR absorption relative to the X-ray
absorption are partly the same as for the case of the optical absorption. However, an
important additional factor contributing IR/X-ray absorption mismatch is that the mid-IR
emission is extended, at least on the pc-scale, i.e. on a scale comparable or larger
than the dense X-ray absorbing medium.

\section{X-ray absorption at high luminosities and at high redshift}

Until a few years ago the existence of obscured QSOs (QSO2s) was under question.
QSO2s are difficult to find because both absorbed and rare. However, recently
a large number of QSO2s have been found in by systemic surveys over large
sky areas, in the optical \citep{zakamska03,ptak06}, in the X-rays
\citep{norman02,fiore03,barger05,silverman05,maccacaro04,maiolino06},
in the radio \citep{donley05,belsole06} and in the infrared \citep{msansigre05,polletta06,
alonsoh06,franceschini05}.

The large number of newly discovered QSO2s has allowed the investigation of
the fraction of obscured AGNs as a function of luminosity. By using the results from 
hard X-ray
surveys, various authors have found evidence for a decreasing fraction of obscured AGNs
with increasing luminosity \citep{ueda03,lafranca05,barger05,akylas06}, although this
result has been questioned \citep{dwelly06}. The same trend was found 
by \cite{simpson05} among optically selected AGNs. If confirmed,
these results
can be interpreted within the scenario of the so-called ``receding torus''
\citep{lawrence91}, where the dependence of the dust sublimation radius with
luminosity  causes the covering factor of the absorbing medium to decrease with
luminosity. Alternatively, \cite{lamastra06} suggested that the dependence of the covering
factor with luminosity is an indirect consequence of the gravitational effects of the black
holes, which is on average more massive in more luminous AGNs, because of selection
effects.

One should keep in mind that, at least in hard X-ray surveys, the census is limited
to the Compton thin sources. Indeed, the faintness of Compton thick AGNs even in the hard
X-rays prevents their detection at cosmological distances \citep[except for a tail of the
population,][]{tozzi06}. To infer the fraction of Compton thick sources at high redshift
we have to rely on other, indirect indicators.
One possibility is to exploit the shape of the hard X-ray background. \cite{gilli06} show
that the population of (Compton thin) AGNs that resolve the X-ray background at
2--10~keV fall short to account for the peak of the latter at 30~keV. The additional
contribution by a population of
Compton thick AGNs with $\rm 10^{24}<N_H<10^{25}~cm^{-2}$ is required to match the shape and
intensity of the X-ray background at energies higher than 10~keV. The required proportion of
Compton thick AGNs with $\rm 10^{24}<N_H<10^{25}~cm^{-2}$ relative to Compton thin AGNs must
be 1:2, i.e. as observed in local AGNs. Note however, that the X-ray background is
insensitive to the population of totally Compton thick AGNs with $\rm N_H > 10^{25}~cm^{-2}$
(since their emission is totally suppressed at any energies),
which therefore remains unconstrained.

An alternative
method to identify Compton thick AGNs at high redshift is by means of mid-IR
data. Indeed, recent Spitzer observations have discovered
a large population of high-z AGNs (identified through a mid-IR, AGN-like excess)
that do not have hard-X counterpart even in deep X-ray observations, and therefore are
likely Compton thick AGNs \citep{alonsoh06,polletta06,donley05}.
Many of the Spitzer studies on high-z AGNs are still ongoing, therefore this field is
currently in continuous evolution. However, results published so far
suggest that the Compton thick AGNs at high-z (including those with $\rm N_H >
10^{25}~cm^{-2}$) are as numerous as Compton thin AGNs, i.e. matching the same relative
fractions observed in the local universe.

\section{Open issues}

The physical and statistical properties of the X-ray absorber in AGNs
are far from being fully understood, and several questions remain unanswered yet.

We have shown that there is clear evidence for a compact absorber, located on the scale
of the BLR, in a few sources. However, it is not clear whether such a compact absorber is
ubiquitous to {\it all} AGNs or not. A related issue is whether {\it all}
AGNs have Compton thick gas around them (including those which are Compton thin along our
line of sight) or not. For what concerns the structure of the absorber, it
is not clear yet whether the temporal variations of $\rm N_H$ are tracing a medium with two
phases (cold, dense clouds inside a warm medium), or a more homogeneous, cold gas with
density gradients. In terms of stability of the X-ray absorber, it is not clear how its
vertical structure (required to account for the large covering factor) is supported.

\begin{SCfigure}
\includegraphics[width=0.5\textwidth]{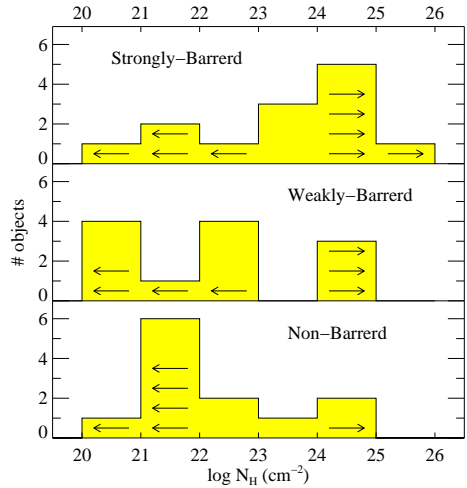}
\caption{Column density distribution of the sources in the samples of \cite{cappi06} and
\cite{guainazzi05}, divided as a function of the strength of the stellar bar in the host galaxy. There is a significant progression
of the average $\rm N_H$ as a function of the bar strength. In particular, most Compton thick
nuclei are hosted in barred galaxies.}
\label{nh_bar}
\end{SCfigure}

Another class of issues is related to some puzzling comparisons of the
X-ray absorption with other observed quantities. \cite{zhang06} found that AGNs with nuclear
$\rm H_2O$ maser disks are not preferentially Compton thick. However, the detection of maser
emission requires large columns of gas ($\rm N_H > 10^{23-24}cm^{-2}$) and very small disk
inclination ($\rm < 10^{\circ}$), the combination of which are expected to produce Compton
thick absorption, that instead is observed for only half of the AGNs in the maser
sample of \cite{zhang06} (i.e. the same fraction observed in
optical samples). On the large scales there is an opposite issue. \cite{maiolino99} found
that the X-ray absorption is correlated with the presence of a bar in the host galaxy. In
particular, Compton thick AGNs appears preferentially hosted in barred galaxies.
As discussed above, Compton thick absorption probably occurs on the sub-parsec scale,
therefore it is hard to understand how the physics of the sub-parsec medium can be
affected by the dynamical properties of the host galaxies on the kpc scale.
We have verified the \cite{maiolino99} result
by using the more recent data by \cite{cappi06} and
\cite{guainazzi05}; but even in these samples the correlation remains, as illustrated in
Fig.\ref{nh_bar}.

\acknowledgements 

This work was supported by the Italian Space Agency.
RM is grateful to the organizers of the conference for their kind invitation.


\begin{thebibliography}{}

\bibitem[Akylas et al.(2006)]{akylas06} Akylas, A., et al.\ 
2006, \aap, 459, 693

\bibitem[Alonso-Herrero et al.(2006)]{alonsoh06} Alonso-Herrero, 
A., et al.\ 2006, \apj, 640, 167

\bibitem[Barger et al.(2005)]{barger05} Barger, A.~J., et al.\ 2005, \aj, 129, 578

\bibitem[Bassani et al.(1999)]{bassani99} Bassani, L., et al.\ 1999, \apjs, 121, 473

\bibitem[Bassani et al.(2006)]{bassani06} Bassani, L., et al.\ 
2006, \apjl, 636, L65

\bibitem[Belsole et al.(2006)]{belsole06} Belsole, E., Worrall, 
D.~M., \& Hardcastle, M.~J.\ 2006, \mnras, 366, 339

\bibitem[Braito et al.(2004)]{braito04} Braito, V., et al.\ 
2004, \aap, 420, 79

\bibitem[Cappi et al.(2006)]{cappi06} Cappi, M., et al.\ 2006, 
\aap, 446, 459

\bibitem[Cocchia et al.(2007)]{cocchia07} Cocchia, F., et al.\ 2007, 
\aap, in press (astro-ph/0612023)

\bibitem[Comastri et al.(2002)]{comastri02} Comastri, A., et al.\ 
2002, \apj, 571, 771

\bibitem[Crenshaw \& Kraemer(1999)]{crenshaw99} Crenshaw, D.~M., 
\& Kraemer, S.~B.\ 1999, \apj, 521, 572

\bibitem[Della Ceca et al.(2002)]{dellaceca02} Della Ceca, R., et 
al.\ 2002, \apjl, 581, L9

\bibitem[Donley et al.(2005)]{donley05} Donley, J.~L., Rieke, 
G.~H., Rigby, J.~R., \& P{\'e}rez-Gonz{\'a}lez, P.~G.\ 2005, \apj, 634, 169

\bibitem[Dwelly \& Page(2006)]{dwelly06} Dwelly, T., \& Page, 
M.~J.\ 2006, \mnras, 372, 1755

\bibitem[Elvis et al.(2004)]{elvis04} Elvis, M., et al.\ 2004, \apjl, 615, 
L25

\bibitem[Evans et al.(2006)]{evans06} Evans, D.~A., et al.\ 2006, \apj, 642, 
96

\bibitem[Fiore et al.(2003)]{fiore03} Fiore, F., et al.\ 2003, 
\aap, 409, 79

\bibitem[Franceschini et al.(2005)]{franceschini05} Franceschini, A., 
et al.\ 2005, \aj, 129, 2074

\bibitem[Gallagher et al.(2006)]{gallagher06} Gallagher, S.~C., 
et al.\ 2006, \apj, 644, 709

\bibitem[Gilli et al.(2000)]{gilli00} Gilli, R., et al.\ 
2000, \aap, 355, 485

\bibitem[Gilli et al.(2006)]{gilli06} Gilli, R., et al. 2006,
\aap, in press (astro-ph/0610939)

\bibitem[Gon{\c c}alves et al.(2006)]{gonclaves06} Gon{\c c}alves, 
A.~C., et al.\ 2006, \aap, 451, L23

\bibitem[Grandi et al.(2006)]{grandi06} Grandi, P., Malaguti, 
G., \& Fiocchi, M.\ 2006, \apj, 642, 113


\bibitem[Guainazzi et al.(2000)]{guainazzi00} Guainazzi, M., et al.\ 2000, \aap, 
356, 463

\bibitem[Guainazzi et al.(2002)]{guainazzi02} Guainazzi, M., Matt, 
G., Fiore, F., \& Perola, G.~C.\ 2002, \aap, 388, 787

\bibitem[Guainazzi et al.(2005)]{guainazzi05} Guainazzi, M., Matt, 
G., \& Perola, G.~C.\ 2005, \aap, 444, 119

\bibitem[Hall et al.(2006)]{hall06} Hall, P.~B., et al.\ 2006, \aj, 132, 1977

\bibitem[Hamann et al.(2002)]{hamann02} Hamann, F., et al.\ 2002, \apj, 564, 592

\bibitem[Hardcastle et al.(2006)]{hardcastle06} Hardcastle, M.~J., 
Evans, D.~A., \& Croston, J.~H.\ 2006, \mnras, 370, 1893

\bibitem[Ho et al.(1997)]{ho97} Ho, L.~C., Filippenko, 
A.~V., \& Sargent, W.~L.~W.\ 1997, \apjs, 112, 315


\bibitem[Imanishi et al.(2006)]{imanishi06} Imanishi, M., Dudley, 
C.~C., \& Maloney, P.~R.\ 2006, \apj, 637, 114

\bibitem[Kaspi et al.(2002)]{kaspi02} Kaspi, S., et al.\ 2002, 
\apj, 574, 643

\bibitem[Krabbe et al.(2001)]{krabbe01} Krabbe, A., B{\"o}ker, 
T., \& Maiolino, R.\ 2001, \apj, 557, 626

\bibitem[Krongold et al.(2003)]{krongold03} Krongold, Y., 
et al.\ 
2003, \apj, 597, 832

\bibitem[La Franca et al.(2005)]{lafranca05} La Franca, F., et 
al.\ 2005, \apj, 635, 864

\bibitem[Lamastra et al.(2006)]{lamastra06} Lamastra, A., Perola, 
G.~C., \& Matt, G.\ 2006, \aap, 449, 551

\bibitem[Lawrence(1991)]{lawrence91} Lawrence, A.\ 1991, \mnras, 
252, 586

\bibitem[Lutz et al.(2004)]{lutz04} Lutz, D., Maiolino, R., 
Spoon, H.~W.~W., \& Moorwood, A.~F.~M.\ 2004, \aap, 418, 465

\bibitem[Maccacaro et al.(2004)]{maccacaro04} Maccacaro, T., 
et al.\ 2004, \apjl, 
617, L33

\bibitem[Maiolino \& Rieke(1995)]{maiolino95} Maiolino, R., \& 
Rieke, G.~H.\ 1995, \apj, 454, 95

\bibitem[Maiolino et al.(1998)]{maiolino98} Maiolino, R., et al.\ 1998, \aap, 338, 781

\bibitem[Maiolino et al.(1999)]{maiolino99} Maiolino, R., 
Risaliti, G., \& Salvati, M.\ 1999, \aap, 341, L35

\bibitem[Maiolino et al.(2001a)]{maiolino01a} Maiolino, R., et al.\ 2001a, \aap, 365, 28

\bibitem[Maiolino et al.(2001b)]{maiolino01b} Maiolino, R., Marconi, 
A., \& Oliva, E.\ 2001b, \aap, 365, 37

\bibitem[Maiolino et al.(2001c)]{maiolino01c} Maiolino, R., 
Mannucci, F., Baffa, C., Gennari, S., \& Oliva, E.\ 2001, \aap, 372, L5

\bibitem[Maiolino et al.(2003)]{maiolino03} Maiolino, R., et al.\ 
2003, \mnras, 344, L59

\bibitem[Maiolino et al.(2006)]{maiolino06} Maiolino, R., et al.\ 
2006, \aap, 445, 457

\bibitem[Malkan et al.(1998)]{malkan98} Malkan, M.~A., Gorjian, 
V., \& Tam, R.\ 1998, \apjs, 117, 25

\bibitem[Markwardt et al.(2005)]{markwardt05} Markwardt, C.~B., 
et al.\ 2005, \apjl, 633, L77

\bibitem[Mart{\'{\i}}nez-Sansigre et al.(2005)]{msansigre05} 
Mart{\'{\i}}nez-Sansigre, A., et al.\ 2005, \nat, 436, 666

\bibitem[Matt et al.(1997)]{matt97} Matt, G., et al.\ 1997, 
\aap, 325, L13

\bibitem[Matt et al.(2003)]{matt03} Matt, G., Guainazzi, M., 
\& Maiolino, R.\ 2003, \mnras, 342, 422

\bibitem[Moran et al.(2002)]{moran02} Moran, E.~C., Filippenko, 
A.~V., \& Chornock, R.\ 2002, \apjl, 579, L71

\bibitem[Nagao et al.(2006a)]{nagao06a} Nagao, T., Marconi, A., 
\& Maiolino, R.\ 2006a, \aap, 447, 157

\bibitem[Nagao et al.(2006b)]{nagao06b} Nagao, T., Maiolino, R., 
\& Marconi, A.\ 2006b, \aap, 447, 863

\bibitem[Netzer et al.(2003)]{netzer03} Netzer, H., et al.\ 
2003, \apj, 599, 933

\bibitem[Norman et al.(2002)]{norman02} Norman, C., et al.\ 
2002, \apj, 571, 218

\bibitem[Polletta et al.(2006)]{polletta06} Polletta, M.~d.~C., et 
al.\ 2006, \apj, 642, 673

\bibitem[Ptak et al.(2006)]{ptak06} Ptak, A., et al.\ 2006, \apj, 637, 147

\bibitem[Reynolds(1997)]{reynolds97} Reynolds, C.~S.\ 1997, 
\mnras, 286, 513

\bibitem[Risaliti et al.(1999)]{risaliti99} Risaliti, G., 
Maiolino, R., \& Salvati, M.\ 1999, \apj, 522, 157

\bibitem[Risaliti et al.(2002)]{risaliti02} Risaliti, G., Elvis, 
M., \& Nicastro, F.\ 2002, \apj, 571, 234

\bibitem[Risaliti et al.(2005)]{risaliti05} Risaliti, G., Elvis, 
M., Fabbiano, G., Baldi, A., \& Zezas, A.\ 2005, \apjl, 623, L93

\bibitem[Risaliti et al.(2006)]{risaliti06} Risaliti, G., et al.\ 
2006, \mnras, 365, 303

\bibitem[Salvati \& Maiolino(2000)]{salvati00} Salvati, M., \& 
Maiolino, R.\ 2000, Large Scale Structure in the X-ray Universe, 
eds.~Plionis, M.~ Georgantopoulos, I., Atlantisciences, Paris, 
France, p.277, 277

\bibitem[Sambruna et al.(1999)]{sambruna99} Sambruna, R.~M., 
Eracleous, M., \& Mushotzky, R.~F.\ 1999, \apj, 526, 60

\bibitem[Schinnerer et al.(1999)]{schinnerer99} Schinnerer, E., 
Eckart, A., \& Tacconi, L.~J.\ 1999, \apjl, 524, L5

\bibitem[Severgnini et al.(2003)]{severgnini03} Severgnini, P., et 
al.\ 2003, \aap, 406, 483

\bibitem[Shi et al.(2006)]{shi06} Shi, Y., et al.\ 2006, 
\apj, in press (astro-ph/0608645)

\bibitem[Silva et al.(2004)]{silva04} Silva, L., Maiolino, R., 
\& Granato, G.~L.\ 2004, \mnras, 355, 973

\bibitem[Silverman et al.(2005)]{silverman05} Silverman, J.~D., et 
al.\ 2005, \apj, 618, 123

\bibitem[Simpson(2005)]{simpson05} Simpson, C.\ 2005, \mnras, 
360, 565

\bibitem[Szokoly et al.(2004)]{szokoly04} Szokoly, G.~P., et al.\ 
2004, \apjs, 155, 271

\bibitem[Tozzi et al.(2006)]{tozzi06} Tozzi, P., et al.\ 2006, 
\aap, 451, 457

\bibitem[Ueda et al.(2003)]{ueda03} Ueda, Y., Akiyama, M., 
Ohta, K., \& Miyaji, T.\ 2003, \apj, 598, 886

\bibitem[Vignati et al.(1999)]{vignati99} Vignati, P., et al.\ 
1999, \aap, 349, L57

\bibitem[Wilkes et al.(2002)]{wilkes02} Wilkes, B.~J., et al.\ 
2002, \apjl, 564, L65

\bibitem[Zakamska et al.(2003)]{zakamska03} Zakamska, N.~L., et 
al.\ 2003, \aj, 126, 2125

\bibitem[Zhang et al.(2006)]{zhang06} Zhang, J.~S., et al.\ 
2006, \aap, 450, 933

\end{thebibliography}
\end{document}